\documentclass{aa}
\usepackage{psfig}
\usepackage{graphicx}
\begin{document}

   \title{On the distance, reddening and progenitor of V838 
          Mon\thanks{Tables~3 and 4 available only in electronic form (ASCII format)
          at CDS via anonymous ftp to cdsarc.u-strasbg.fr (130.79.128.5) or via
          http://cdsweb.u-strasbg.fr/cgi-bin/qcat?J/A+A/}}

   \author{
           U.Munari\inst{1}
           \and
           A.Henden\inst{2}
           \and
	   A.Vallenari\inst{3}
           \and
           H.E.Bond\inst{4}
	   \and
	   R.L.M.Corradi\inst{5}
	   \and
           L.Crause\inst{6}
	   \and
           S.Desidera\inst{3}
           \and
	   E.Giro\inst{3}
	   \and
	   P.M.Marrese\inst{1}
	   \and
           S.Ragaini\inst{3}
	   \and
           A.Siviero\inst{1}
	   \and
           R.Sordo\inst{1}
	   \and
	   S.Starrfield\inst{7}
           \and
	   T.Tomov\inst{8}
	   \and
	   S.Villanova\inst{9}
	   \and
	   T.Zwitter\inst{10}
	   \and
           R.M.Wagner\inst{11}
          }

   \offprints{U.Munari (munari@pd.astro.it)}

   \institute{
INAF-Osservatorio Astronomico di Padova, Sede di Asiago,
I-36012 Asiago (VI), Italy
	\and
Univ. Space Research Assoc./U. S. Naval Observatory,
P. O. Box 1149, Flagstaff AZ 86002-1149, USA
	\and
INAF-Osservatorio Astronomico di Padova, Vicolo dell'Osservatorio 8,
35122 Padova, Italy
	\and
Space Telescope Science Institute, 3700 San Martin Drive, Baltimore, 
MD 21218  USA
	\and
Isaac Newton Group of Telescopes, Apartado de Correos 321, 38700 Santa
Cruz de La Palma, Canarias, Spain
	\and
South African Astronomical Observatory, P.O.Box 9, Observatory 7935, 
South Africa
	\and
Dept. of Physics and Astronomy, Arizona State Univ., P. O. Box 871504 Tempe,
AZ 85287-1504, USA
        \and 
Centre for Astronomy, Nicolaus Copernicus University, ul. Gagarina 11, 
87-100 Torun, Poland
	\and
Dipartimento di Astronomia, Universit\'{a} di Padova, 35122 Padova, Italy
	\and
University of Ljubljana, Department of Physics, Jadranska 19, 
1000 Ljubljana, Slovenia
	\and
Large Binocular Telescope Observatory, Univ. of Arizona, 933 North
Cherry Avenue, Tucson, AZ 85721, USA
        }

   \date{Received ... / Accepted ...}

\abstract{Extensive optical and infrared photometry as well as 
low and high resolution spectroscopy are used as inputs in deriving robust
estimates of the reddening, distance and nature of the progenitor of
V838~Mon, the 2002 outbursting event that produced a most spectacular
light-echo. The reddening affecting V838~Mon is found to obey
the $R_V$=3.1 law and amounts to ($i$) $E_{B-V}$=0.86 from the interstellar NaI
and KI lines, ($ii$) $E_{B-V}$=0.88 from the energy distribution of the B3\,V
component and ($iii$) $E_{B-V}$=0.87 from the progression of extinction along the
line of sight. The adopted $E_{B-V}$=0.87$\pm$0.01 is also the amount
required by fitting the progenitor with theoretical isochrones of
appropriate metallicity. The distance is estimated from ($a$) the galactic
kinematics of the three components of the interstellar lines, ($b$) the
amount of extinction vs the HI column density and vs the dust emission
through the whole Galaxy in that direction, from ($c$) spectrophotometric
parallax to the B3\,V companion, from ($d$) comparison of the observed
color-magnitude diagram of field stars with 3D stellar population models of
the Galaxy, from ($e$) comparison of theoretical isochrones with the
components of the binary system in quiescence and found to be around 
10~kpc.
Pre-outburst optical and IR energy distributions show that the component
erupting in 2002 was brighter and hotter than the B3\,V companion. The best
fit is obtained for a 50\,000~K source, 0.5 mag brighter than the B3\,V
companion. The latter passed unaffected through the outburst, which implies
an orbital separation wide enough to avoid mass exchange during the
evolution of the binary system, and to allow a safe comparison with
theoretical isochrones for single stars. Such a comparison suggests that the
progenitor of the outbursting component had an initial mass
$\sim$65~M$_\odot$, that it was approaching the Carbon ignition stage in its
core at the time it erupted in 2002 and that the age of the V838~Mon binary
system is close to 4 million yr. The 2002 event is probably just a shell
thermonuclear event in the outer envelope of the star.

\keywords{Stars: evolution - Stars: early type - Stars: individual: 
           V838 Mon - Stars: winds, outflows - ISM: dust,extinction -- ISM: kinematics}
          }

   \maketitle

\section{Introduction}

   \begin{figure}[t]
   \centerline{\psfig{file=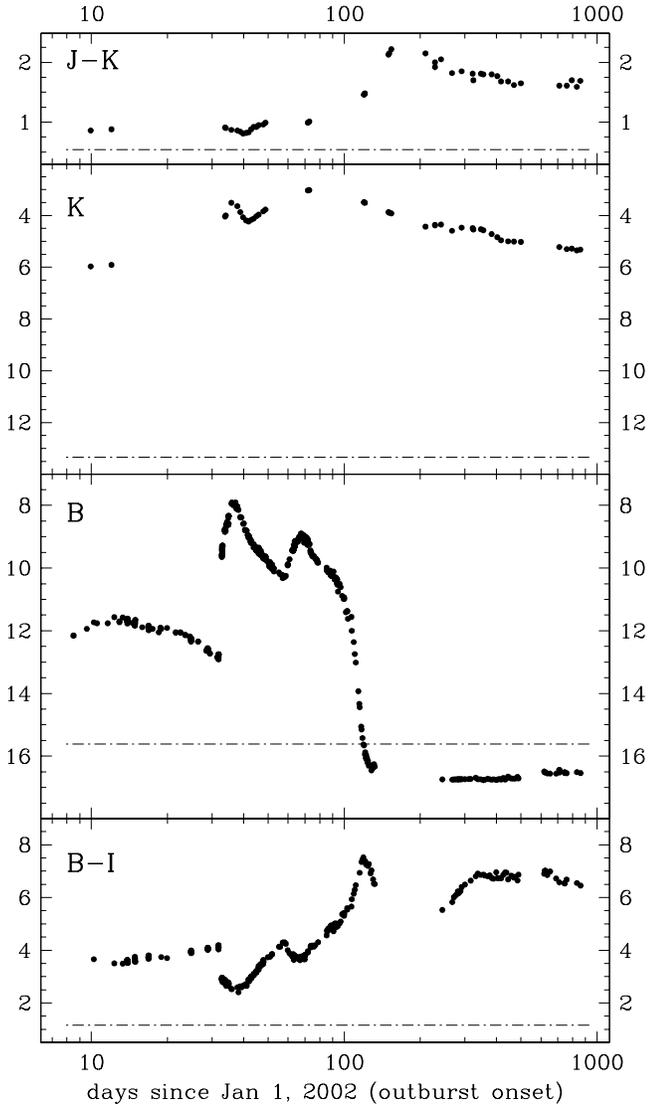,width=8.6cm}}
   \caption[]{Optical and IR light-curves of the outburst of V838~Mon as
   recorded with USNO 1.0m and 1.55m telescopes; some early-time observations
   come from other sources as identified in Munari et al. (2002b).
   The outburst onset is
   around Jan 1, 2002, and the first and last observations in the picture
   are for the dates Jan 9, 2002 and May 11, 2004, respectively. 
   The gaps in the optical
   light-curves correspond to seasonal invisibility. The dotted-dashed lines
   give the values for the quiescence prior to the outburst.}
   \label{lightcurve}
   \end{figure}

V838~Mon rose from obscurity in early January 2002, when it was discovered in
outburst by Brown (2002). The unusually cool spectrum and strange light-curve
helped to keep attention focused on the object for the next three 
months, until the discovery in late March by Henden, Munari and Schwartz
(2002) of a light-echo rapidly developing around V838~Mon. The presence of
the first Galactic light echo in $\sim$70 years fostered a massive,
multi-wavelength observing campaign for V838~Mon. A high spatial resolution
imaging series of the light-echo expansion and evolution was
collected with HST by Bond et al. (2003), recently expanded by new
images secured within the Hubble Heritage Program. An account of the
spectroscopic, photometric and polarimetric evolution of V838~Mon during the
first season of visibility was presented by Munari et al. (2002a). A major
observational constraint was the discovery by Desidera \& Munari (2002) and
Munari et al. (2002b) that V838~Mon is a binary system
containing a normal B3V star, implying a young age and M$\geq$7~M$_\odot$
for its outbursting companion.  Over the last two years,
V838~Mon has been the topic of numerous refereed papers and conference reports,
with a sample given below.
IR photometry and spectroscopy was presented and discussed
by Banerjee \& Ashok (2002), Crause et al. (2003), Evans et al. (2003) and
Rushton et al. (2003), optical polarimetry and spectropolarimetry by
Wisniewski et al. (2003a,b) and Desidera et al. (2004), optical photometry by
Kimeswenger et al. (2002), Barsukova et al. (2002), Kato (2003) and Goranskii
et al. (2004), and optical spectroscopy by Munari et al. (2002a,c),
Goranskii et al. (2002), Kolev et al. (2002), Osiwala et al. (2002), Kipper
et al. (2004). Modeling of the outburst was presented by Retter \&
Marom (2003), Soker \& Tylenda (2003) and Boschi \& Munari (2004), while
modeling of the light-echo expansion was performed by Bond et al.
(2003), Sugerman (2003) and Tylenda (2004).

   \begin{table}[t]
   \caption[]{Open clusters within 5$^\circ$ of V838~Mon from the Lyng\aa\ (1987)
   catalog. $\theta$ is the angular distance in degrees from V838~Mon.
   Reddening and distances are taken from the Dias et al. (2002)
   compilation.}
   \centerline{\psfig{file=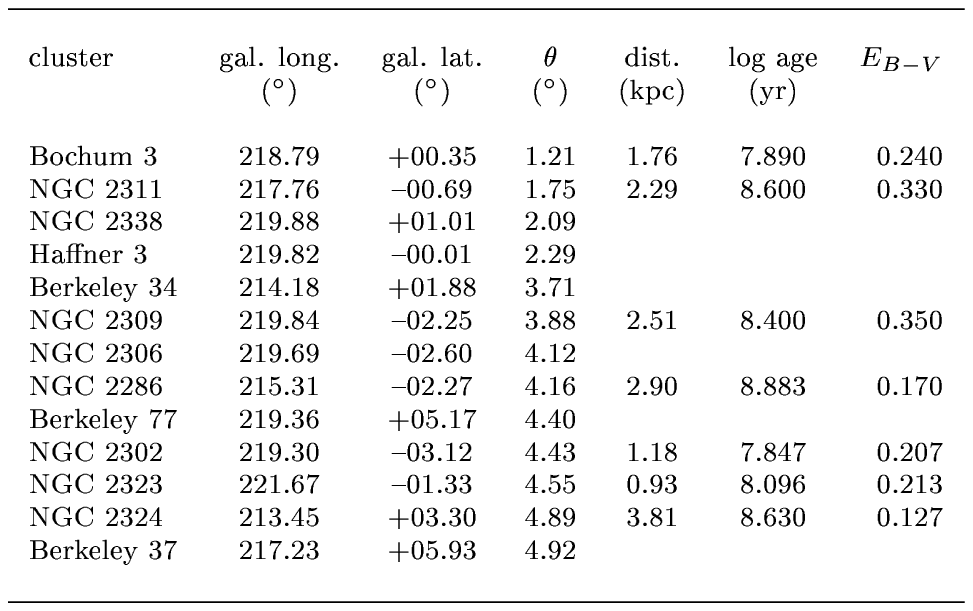,width=8.8cm}}
   \end{table}
   \begin{figure}[t]
   \centerline{\psfig{file=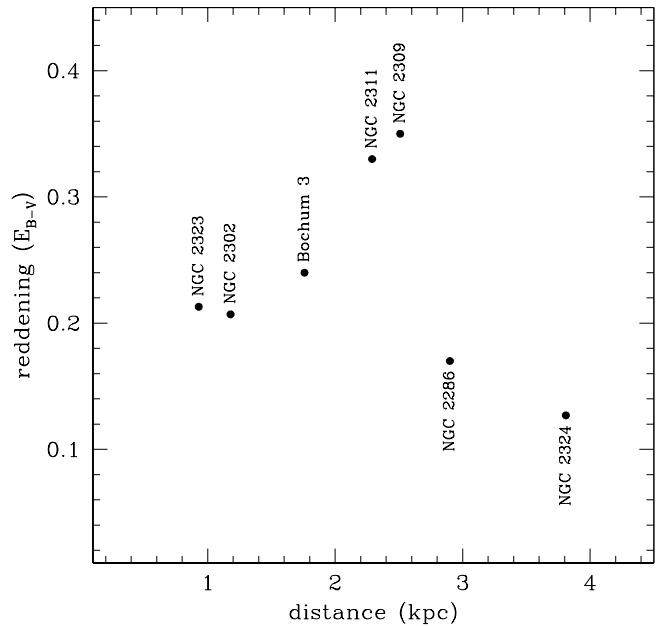,width=8.6cm}}
   \caption[]{Extinction as function of distance for open clusters within
   5$^\circ$ from V838 Mon.}
   \label{clusters}
   \end{figure}

Common to many of these papers were the issues of reddening, distance and
nature of the progenitor of V838~Mon. However, none of these papers focused
specifically on these topics, and limited their discussion to the derivation
or adoption of approximate values based on {\em assumed} energy distribution
of the outbursting component. The aim of the present paper is to attempt an
accurate derivation of reddening and distance of V838~Mon by careful
comparison of different techniques and approaches that do not involve the
outbursting component, as well as to derive by comparison with theoretical
models the nature and evolutionary state of V838~Mon prior to the onset of
the outburst.
 
Optical and IR photometry used in this paper was obtained with CCD-equipped
USNO Flagstaff Station (NOFS) telescopes. The {\sl UBV$R_{\rm C}$I$_{\rm
C}$} data were obtained with the 1.0m telescope, the JHK$^\prime$L$^\prime$
with the 1.55m telescope.  The optical photometry is strictly tied to the
Landolt (1983, 1992) system of equatorial standards, while the infrared
photometry uses standard JHK' filters and differential measures with
respect to local 2MASS stars. The low resolution optical spectrophotometry
of V838~Mon was obtained with the AFOSC+CCD imager+spectrograph of the 1.82m
telescope operated in Asiago by the Astronomical Observatory of Padova. At
the same telescope we secured high resolution spectra with the Echelle+CCD
spectrograph. Other high resolution spectra of V838~Mon were obtained with
the Coude spectrograph+CCD of the Rozhen 2m telescope, with GIRAFFE at the
1.93m SAAO telescope, with SARG at the 3.5m TNG telescope, and with FEROS at
the 2.2m ESO telescope in La Silla. Further details will be provided below
where necessary.

Figure~1 presents a review of the optical and IR photometric evolution of
V838~Mon, updated to the end of the third season of visibility in May 2004.
With the exception of some early-time datapoints, the light-curves are
entirely based on data from the USNO telescopes and therefore they are
highly consistent and free from systematic effects of color transformation
between different local photometric systems that would badly affect
observations from different observatories of an object with such extreme
colors as V838~Mon.

\section{Evidence of patchy extinction}

V838~Mon lies close to galactic equator ($b$=+1$^\circ$) and in the
anti-center quadrant ($l$=218$^\circ$). I spite the line of sight crosses
the Orion, Perseus and Outer spiral arms, inspection of the Palomar plates,
of the 2MASS, IRAS, UKST-H$\alpha$ maps and the Neckel and Klare (1980)
extinction charts do support an apparently smooth star counts distribution
for several degrees around V838~Mon position. Therefore in principle, open
clusters and field stars with accurate photometry and spectroscopic
classification can be used to search for a relationship between reddening
and distance in the direction of V838~Mon.

Within 5$^\circ$ from V838~Mon there are 13 open clusters in the
catalog of Lyng\aa\ (1987). They are listed in Table~1. According to the
compilation of Dias et al. (2002), only half of them have published
determinations of distance and reddening. The placing of these 7 clusters on
a distance-reddening diagram (cf. Figure~2) shows a large scatter and
basically no obvious trend over such angular distances.

The best source of homogeneous spectroscopic classification of field stars is
the Michigan Project, which is a continuing program at the University of
Michigan where all HD stars are being reclassified on the MK system,
starting at the south Galactic pole, and ending at the north Galactic pole.
Volume 5 (Houk \& Swift 1999) covers the declination strip $-12^\circ \leq \delta \leq
5^\circ$, thus including the region around V838~Mon.

We selected all HD stars within a radius of 3$^\circ$ from V838~Mon and
retrieved their $B_{\rm T}$ and $V_{\rm T}$ magnitudes from the Tycho-2
catalog (H\o g et al. 2000). The latter were transformed into corresponding
Johnson $B_{\rm J}$ and $V_{\rm J}$ following Bessell (2000) transformations.
The reddening of each star was than derived by comparison with Fitzgerald
(1970) intrinsic colors, and the distance computed using the absolute
magnitudes of the Michigan Project scale (from N. Houk web
page\footnote{http://www.astro.lsa.umich.edu/users/hdproj/\\
mosaicinfo/absmag.html}).

   \begin{figure}[t]
   \centerline{\psfig{file=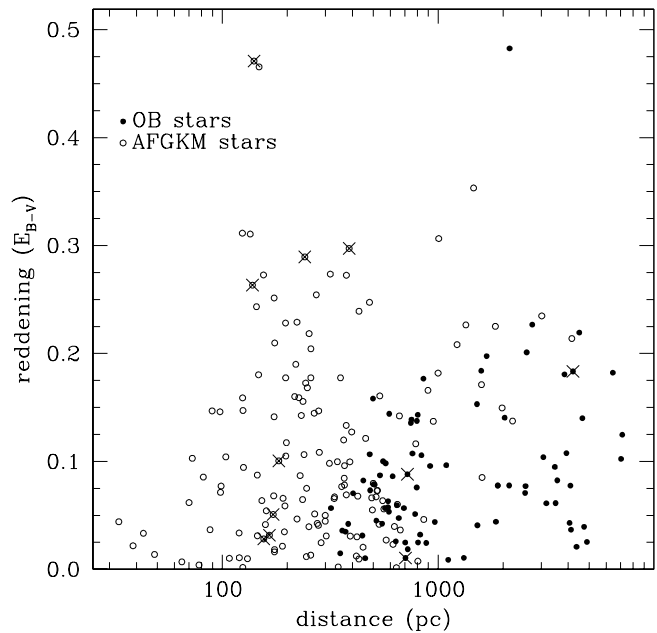,width=8.6cm}}
   \caption[]{The HD stars closer than 3$^\circ$ from V838~Mon are plotted.
   The crosses mark those within 0.6$^\circ$ from V838~Mon. Only HD stars
   classified by the Michigan Project and with Tycho-2 $B_{\rm T}$,$V_{\rm T}$
   data are considered.}
   \label{fieldHD}
   \end{figure}

Figure~3 plots the distribution of HD field stars in a distance-reddening
diagram. No well defined progression of the reddening with distance is
evident in the figure. The amount of scatter on the diagram is much larger
than justifiable in terms of measurement and classification errors, and it
clearly confirms the evidence from open clusters that the reddening is very
patchy over the sky area centered on V838~Mon in spite of the smooth
star counts distribution over the area. Even restricting to the eleven HD
stars closer than 0.6$^\circ$ (marked by crosses in Figure~3) does not help
to define a well-behaved relation between distance and reddening. This
indicates that the reddening coherence area around V838~Mon has a
significantly smaller radius.

   \begin{figure*}[t]
   \centerline{\psfig{file=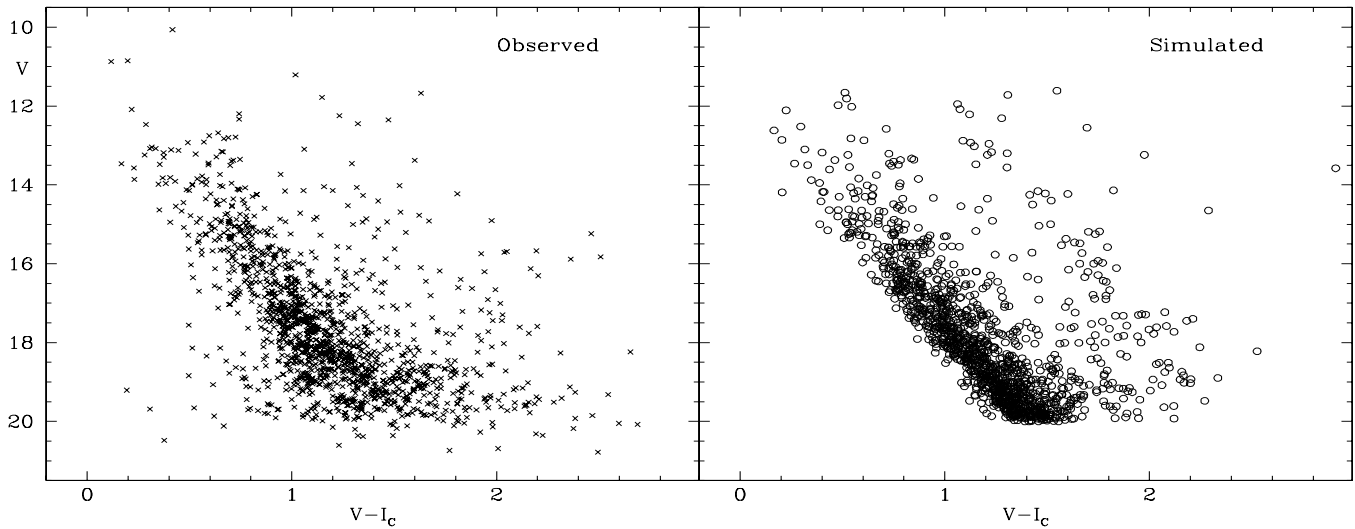,width=18.0cm,height=7.0cm,angle=270}}
   \caption{The left panel plots on a $V$, $V-I$ diagram the stars 
   within a 11$\times$11 arcmin field centered on V838~Mon as observed
   with the NOFS 1.0m telescope. The right panel represents the fit to
   the observed distribution using the Padova Galaxy Model, normalized to
   the same number of stars (1358).}
   \label{CDM}
   \end{figure*}

\section{The field stars close to V838 Mon}

Noting the patchy reddening structure over degree-sized areas, we have
instead focused on a much smaller angular scale in looking for a coherent
relation between distance and reddening along the line of sight to V838~Mon.
Deep {\em UBVR$_{\rm C}$I$_{\rm C}$} photometry of field stars within a 11$\times$11
arcmin field centered on V838~Mon has been obtained with the NOFS 1.0m
telescope. The observed color-magnitude diagrams of these field stars have
then been compared to and fitted with synthetic diagrams from stellar
population models of our Galaxy. The photometric data of the field stars are
electronically available\footnote{At http://ulisse.pd.astro.it/V838\_Mon/}.

   \begin{table}[t]
   \caption[]{Absorption along the line of sight to V838~Mon according to the
   analysis of the color magnitude diagram of the field stars presented in
   Figures~4 and 5.}
   \begin{tabular}{ c c c c c c c c c}
   \hline
   \multicolumn{8}{c}{}\\
   dist (kpc) & 0.1& 1.0&2.5&3.0&4.0& 5.0& 7.0& 10.0 \\
   A$_{\rm V}$ & 0.01& 0.15&0.3&0.7& 1.4& 1.6& 2.1&2.7\\
   \multicolumn{8}{c}{}\\
   \hline
   \end{tabular}
   \end{table}
   \begin{figure}[t]
   \centerline{\psfig{file=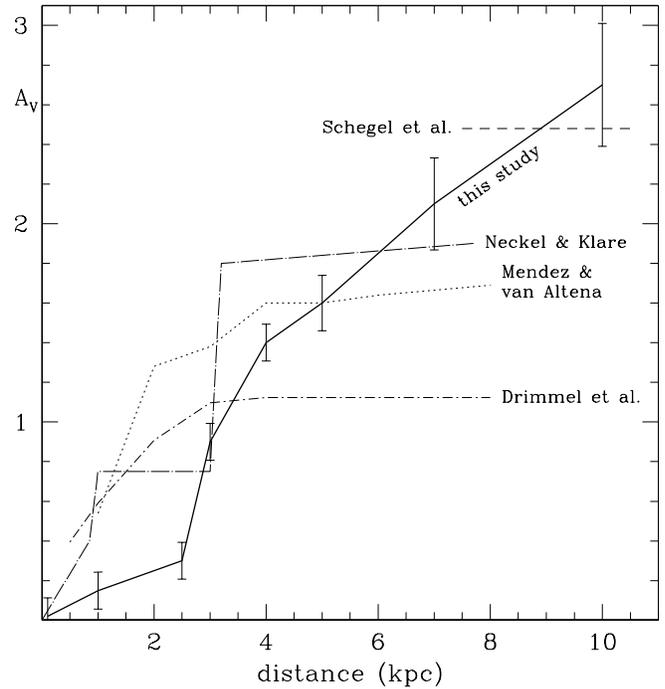,width=8.6cm}}
   \caption[]{Extinction along the line of sight to V838~Mon (thick and solid
   line) as implied by the galactic synthetic population fit in Figure~4 to
   the observed color-magnitude diagram of field stars.  The dotted line
   represents the Mendez \& van Altena (1998) model for the dust in the
   Galaxy, the short dashed-dotted line is the Drimmel et al. (2003) model,
   while the dashed lines give the total galactic extinction along the line
   of sight according to the Schlegel et al (1998) COBE maps. The long
   dashed-dotted line is the extinction behavior according to the Neckel
   \& Klare (1980) maps (poorly populated at $>$3 kpc).}
   \label{Av}
   \end{figure}

V838~Mon lies close to the galactic equator and in the general anti-center
galactic direction, where stellar counts are largely dominated by the disk
population.  Bertelli et al. (1995) showed that the slope of the main
sequence of the disk population in color-magnitude diagrams is mainly
governed by the extinction along the line of sight. At any $V$ magnitude the
bluest stars on the main sequence can be interpreted as the envelope of the
main sequence turnoffs of the population having absolute magnitude
$M_{tur}$, shifted toward fainter magnitudes and redder colors by the
increasing distance and corresponding extinction. Starting from an initial
guess, the amount of extinction at increasing distances is adjusted (via
$\chi ^2$ test on the color distribution as function of magnitude) until a
satisfactory agreement between the main sequence blue edge location in the
observations and in the model is reached. The distribution of observational
errors is evaluated and taken into account in the process.

The Galaxy model that we used is based on the code described by Bertelli et
al. (1995) and recently revised by Vallenari et al. (2000, 2004).  The 
stars are generated according to the evolutionary model of the Galaxy, and
then are distributed along the line of sight following the spatial
model of the Galaxy.  All stellar populations are taken into account.  The
generation of the synthetic population makes use of the set of the stellar
tracks by Girardi et al. (2000).  A double exponential mass distribution
inside the thin disk is adopted having a scale length of 2500 pc and a scale
height varying with the age of the population, ending at 350 pc for the
oldest component. A thick disk component is included with a scale height of
800 pc. A stochastic age metallicity relation for the disk is assumed, with
Z going from 0.008 to 0.03.

We performed the analysis in both $V$,{\em B--V} and $V$,{\em V--I} planes,
with similar results.  Figure~4 presents the distribution of observed and
synthetic stars in the $V$,{\em V--I} diagram, and Figure~5 gives the
resulting distribution of the extinction ($A_{\rm V}$) along the line of
sight. We experimented with different star formation histories in both the
thin and thick galactic disks, but $A_{\rm V}$=$f(d)$ never changed by more
than 0.1 mag at any point along the line of sight. The values reported in
Table~4 and adopted here are those providing the best results in the $\chi
^2$ tests. The error bars in Figure~5 include both the effect of
observational errors and incompleteness as well as different star formation
histories in the galactic disk over the last 10 Gyr. The magnitude of
turnoff stars beyond 6~kpc is at $V\geq$19, where observational errors
become significant and completeness begins to be affected. Therefore, the
results in Table~4 and Figure~5 are less accurate beyond 6~kpc. To
increase the accuracy at larger distances it is advisable that much deeper
photometry is obtained in future observations with 8m class telescopes. 
Comparing the HR diagrams of field stars in different colors it is clear
that a standard $R_{\rm V} = A_{\rm V}/E_{B-V} = $3.1 extinction law
applies.

In Figure~5 comparison is made with other $A_{\rm V}$=$f(d)$ already
presented in literature.  Mendez \& van Altena (1998) make use of the
large-scale properties of the dust layer in the Galaxy to derive the
absorption in the plane of the Galaxy. Their model indicates that a
differential optical absorption of 0.5~mag~kpc$^{-1}$ is adequate to
reproduce the available reddening maps in the range 2$\leq $r$\leq$6~kpc.
The internal uncertainties reported by Mendez \& van Altena are of the order
of $E_{B-V}$=0.25, or $A_V$=0.77 mag for a standard $A_V$=3.1$\times E_{B-V}$. 
Schlegel et al (1998), using the COBE-DIRBE 100 and 240 $\mu$m
all sky survey, derived a map of dust emissivity, which was converted into a
map proportional to dust column density and, finally, to a map of the total
amount of reddening through any direction in the Galaxy. These maps are
expected to suffer in accuracy close to the galactic plane, at $ |b| <$ 5$^\circ$. 
Drimmel et al (2003) recently presented a dimensional model of
the dust distribution based on COBE-DIRBE infrared data which aims to be
useful closer to the Galactic plane, since it accounts for the presence of
spiral arms. As stated by Drimmel et al., regions having anomalous
emission due to warm dust (like Orion or the Galactic center) are not well
described by their model. Neckel \& Klare (1980) extinction maps are based
on observed $B-V$ colors of stars of known spectral type and luminosity
class so that a spectro-photometric parallax can be estimated. Neckel \&
Klare data are only tentative at distances larger than 3~kpc given the
paucity of data.

\section{The NaI and KI interstellar lines}

The interstellar lines are a powerful mean for deriving the
reddening along the line of sight. Munari \& Zwitter (1997) have calibrated
accurate relations between reddening and equivalent width of NaI and KI
interstellar lines that we apply to V838~Mon in this section.

We have secured a number of high resolution spectra during the outburst of
V838~Mon which are suitable to reveal and measure the interstellar NaI and
KI lines. They are summarized in Table~3 (available in electronic form
only). The general appearance of interstellar lines in V838~Mon spectra is
illustrated in Figure~6. There are three main components contributing to the
profiles of interstellar lines. They are seen well separated in the KI line
observed at a resolving power 48\,000 with FEROS at the 2.2m at ESO. The two
components at +29 and +42 km~sec$^{-1}$ are blended at the lower
resolution of the other instruments, particularly in the case of the
stronger NaI lines. For this reason these lower-resolution spectra show a narrow
{\em red component}, at +65.1 km~sec$^{-1}$ radial velocity, and a broader
{\em blue component}, at +37.7 km~sec$^{-1}$ radial velocity. It is worth
noting that the average velocity of the two blue KI components (at +29 and +42
km~sec$^{-1}$) in the FEROS spectrum, weighted according to their equivalent
widths, matches exactly the velocity of the blended blue component (+37.7
km~sec$^{-1}$) in the lower resolution spectra of both NaI and KI lines.

The results of the measurement of heliocentric wavelength, equivalent width
and FWHM of the NaI and KI lines of spectra listed in Table~3 are given in
Table~4 (available in electronic form only), separately for the blue and the
red components. Their mean values and standard deviation are given in
Table~5.  Extensive tests on the spectra suggests that the minimal apparent
changes with time traceable in Table~4 {\em are not} the result of intrinsic
variability of the interstellar lines themselves. Instead, they arise
from the difficulty (arbitrary) in tracing the level of the underlying
{\em continuum} in the continuously changing shape of the underlying broad
stellar P-Cyg profiles (cf. Figure~6), as well as the large assortment of
spectrographs - each with its own PSF - contributing to the monitoring
of the interstellar lines.

   \begin{figure}[t]
   \centering
   \includegraphics[width=8.6cm]{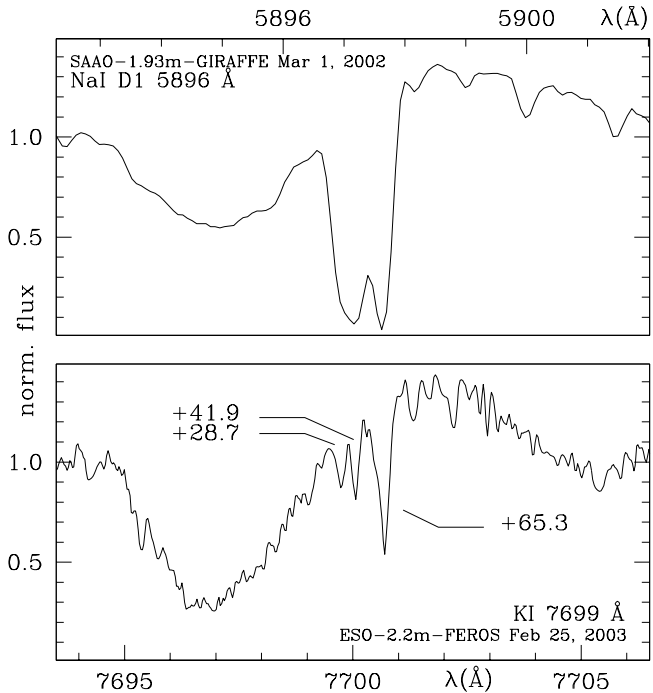}
   \caption{Interstellar lines in the high resolution optical spectra of
   V838~Mon.  The upper panel shows the profile of the NaI~D1 line
   superimposed on the broad P-Cyg stellar line. The profile is composed of
   a narrow red component and a blue unresolved blend. The blue blend is
   resolved in a higher resolution FEROS observations of the KI 7699~\AA\
   line of V838 Mon on the lower panel.}
   \label{KI}
   \end{figure}

   \begin{table}
   \centering
   \caption[]{Journal of the high-resolution spectroscopic observations
   with the Asiago, SAAO, ESO, TNG and Rozhen telescopes (available in 
   electronic form only).}
   \label{Tab_1}
   \end{table}
   \begin{table}
   \centering
   \caption[]{Results of the multi-Gaussian fitting to the NaI and KI
   interstellar lines in the high resolution spectra of V838~Mon listed in
   Table~1 (available in electronic form only).}
   \end{table}

The time behavior of heliocentric radial velocity, equivalent width and
FWHM is plotted in Figure~7. It clearly indicates that since earliest
outburst phases these values have remained constant, being unaffected by the
large changes in the mass loss (amount and velocity) of V838~Mon during its
active state. This indicates that the ejected material had not reached and
swept away any pre-existing circumstellar material giving rise to NaI and KI
absorptions, which is consistent with the absence of X-ray detection in
Chandra observation of V838 Mon by Orio et al. 2003 and the central
dust-free region described by Tylenda (2004).  All three components of the
interstellar lines seem truly interstellar and not circumstellar because
($a$) ground-based and HST images of the light-echo (Munari et al. 2002c,
Bond et al. 2003, and a recent Hubble Heritage Program image) show a hole
surrounding the central star, ($b$) the reddening from interstellar 
lines (cf next section) matches that from field stars, and above all because ($c$)
the radial velocities of the three components of the interstellar lines
(65, 42 and 29 km~sec$^{-1}$) precisely correspond to those of the HI
complexes (68, 43 and 27 km~sec$^{-1}$) observed along the line of sight to
V838~Mon in the Leiden/Dwingeloo HI Survey (Hartmann \& Burton 1997).

   \begin{figure}[t]
   \centering
   \includegraphics[width=8.6cm,height=10.0cm]{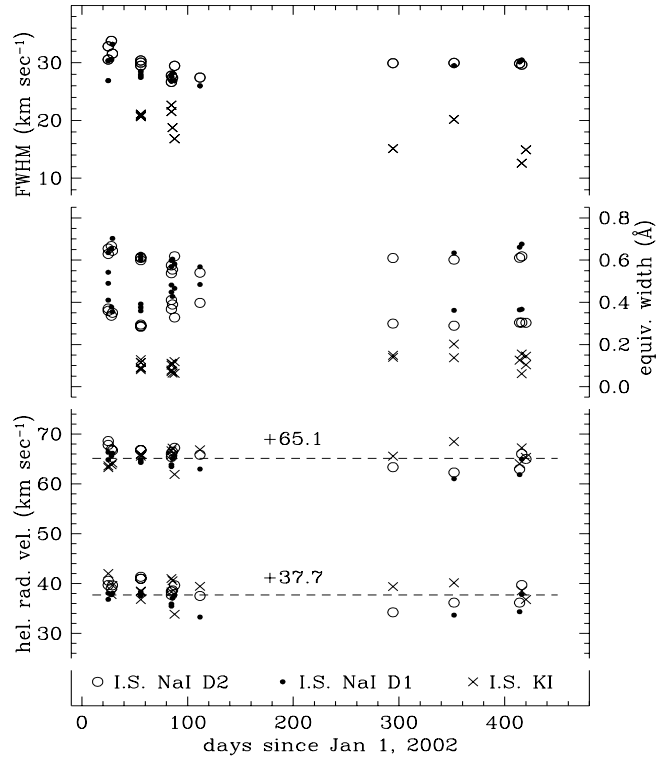}
   \caption{Heliocentric radial velocities, equivalent widths and full width
   at half maximum from Table~4 of the interstellar lines as function of time since the
   onset of the outburst of V838~Mon around Jan 1, 2002. Their 
   constancy suggests that the violent eruption has not interfered with
   possible pre-existing circumstellar material.}
   \label{RV}
   \end{figure}

\subsection{Reddening from interstellar lines}

From the wavelength of maximum polarization, Desidera et al. (2004) have
demonstrated how the reddening affecting V838~Mon follows the normal $R_{\rm
V} = A_{\rm V}/E_{B-V} = $3.1 extinction law. The same result follows from
the analysis of {\em UBVRI} photometry of field stars in sect. 3. We can
therefore safely apply to interstellar lines the Munari \& Zwitter (1997)
calibration for the average interstellar medium to the equivalent width of
NaI and KI lines in Table~5. The resulting reddening is $E_{B-V}$=0.86 from
KI and $E_{B-V}$=0.85 from NaI. The results are in excellent agreement
between the two sets of lines. We adopt $E_{B-V}$=0.86 as the reddening
indicated by the interstellar lines, giving higher weight to the results
from KI lines which are still far from saturation compared to the nearly
saturated NaI lines.

   \begin{table}[t]
   \centering
   \caption[]{Mean values and standard deviations of the radial velocity
   (RV), full width at half maximum ($\Gamma$) and equivalent width (E.W.)
   of the NaI~D and KI interstellar lines in the spectra of V838~Mon
   detailed in Table~2.}
   \centerline{\psfig{file=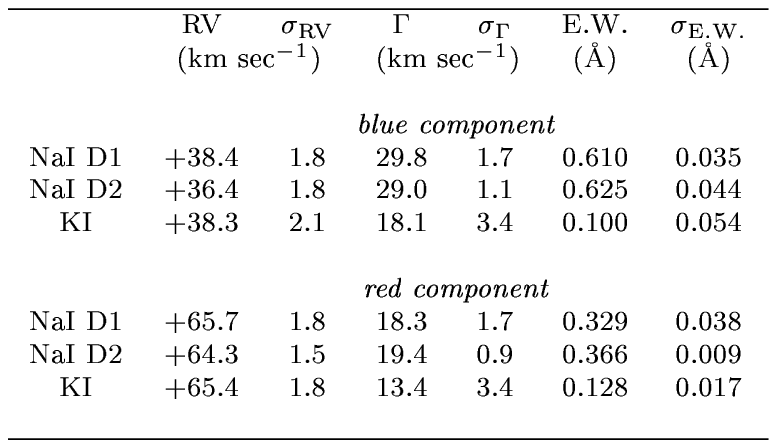,width=7.5 cm}}
   \label{Tab_3}
   \end{table}

\subsection{Distance from radial velocity of interstellar lines and the
            galactic rotation}

Some constraints on the distance to V838~Mon can be derived by combining the
observed radial velocity of interstellar material and the rotation curve of
the Galaxy.

The three principal components of the interstellar absorption lines have
heliocentric radial velocities of +65.1, +41.9 and +28.7 km~sec$^{-1}$. In
Figure~8 they are over-plotted onto the rotation curve of the Galaxy from
Hron (1987) and onto the velocity field in the direction of V838~Mon as
derived by Brand \& Blitz (1993).
 
The velocity of the two components at +41.9 and +28.7 km~sec$^{-1}$ is
compatible with a broad range of distances, while the component at +65.1
km~sec$^{-1}$ supports a distance to the absorbing medium not shorter than
3~kpc, with a corresponding larger distance for V838~Mon itself. Similar
lower limits to the distance to V838~Mon were derived by Wisniewski et al.
(2003a) and Kipper et al. (2004) from similar kinematic arguments based on
the observed velocity of interstellar lines.

\section{The B3V companion}

Munari et al. (2002c) commented on the fact that the {\em UBVR$_{\rm
C}$I$_{\rm C}$} light-curves of V838~Mon started to exhibit a puzzling {\em
bluing} in $U-B$ 90 days and in $B-V$ 115 days after the outburst onset,
while $V-I$ was rising to very red values, approaching those of brown
dwarfs. The reason became apparent when Desidera and Munari (2002)
discovered on spectra at day 274 that the contribution from the very cool
outbursting component was less in the blue part of the spectrum, revealing
the presence of a fainter and hotter companion. V838~Mon was thus shown to be
a binary system, and the hot companion was classified by Munari et al.
(2002b) on spectra for day 300 as a B3\,V star. The presence of the hot
companion was confirmed on higher resolution spectra by Wagner and
Starrfield (2002). The composite nature of optical spectra of V838~Mon
during this phase is illustrated in the top panel of Figure~9. Later on
there was an increase in the contribution of the outbursting component at
shorter wavelengths, as illustrated in the bottom panel of Figure~9.

The average color of V838~Mon at the time of the naked visibility of the
B3\,V companion was {\em B--V}=+0.68 (cf. Figure~1 and Munari et al. 2002b).
Comparing with the intrinsic colors of a B3\,V from Fitzgerald (1970) it
results in $E_{B-V}$=0.88, in excellent agreement with the results from
interstellar NaI and KI lines.

The metallicity in the galactic disk at the galactocentric distance of
V838~Mon is [Fe/H]=$-$0.7, about half dex lower than in the solar
neighborhood (e.g. Friel et al. 2002, Hou et al. 2003).
The effect on the {\em B--V} color of the B3\,V component in V838~Mon of
such a reduction in metallicity is a very minor one, being on the
Rayleigh-Jeans tail of the energy distribution of a hot star. Integrating
the transmission profiles of the Landolt's $B$ and $V$ bands (to which our
photometry is tied) to the 2500-10500~\AA\ synthetic Kurucz spectral
library of Munari et al. (2004), the net effect is just 0.007 mag.

The average $V$ magnitude of V838~Mon at the time the $V$ band was dominated by
the radiation from the B3\,V component is $V$=16.05$\pm$0.05. Coupled with
the $E_{B-V}$=0.87$\pm$0.01 and an absolute magnitude $M_V$=$-$1.70$\pm$0.05
for a B3\,V star from Houk (2004), it implies a distance of 10~kpc to
V838~Mon.

\section{Combined results}

   \begin{figure}
   \centerline{\psfig{file=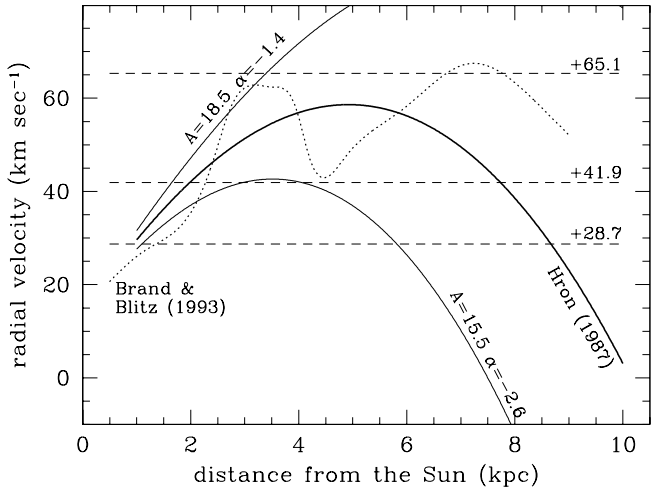,width=8.6cm}}
   \caption[]{The radial velocities of the three components of NaI and KI
    interstellar lines (+28.7, +41.9 and 65.1 km~sec$^{-1}$) are
    over-plotted on the velocity field from Brand \& Blitz (1993) and on
    the rotation curve of the Galaxy of Hron (1987). For the latter
    two additional curves are plotted, corresponding to the extremes of the
    allowed ranges for $A$ and $\alpha$ parameters.}
   \label{Av}
   \end{figure}

	\subsection{Reddening}

Two assumption-independent methods, the interstellar atomic absorption lines
and the colors of the B3V companion, provide consistent results on
the reddening affecting V838~Mon: $E_{B-V}$=0.86 and $E_{B-V}$=0.88,
respectively. There is no differential extinction between the outbursting
star and the B3V companion.

The same amount of reddening is derived from the modeling of the HR diagram
of field stars close to V838~Mon performed in sect. 3. In fact, entering
Table~2 or Figure~5 with the distance (10~kpc) to V838~Mon from the
spectro-photometric parallax to the B3V companion, $E_{B-V}$=0.87 is
obtained.

We therefore conclude that the extinction toward V838~Mon follows the
standard $R_{\rm V} = A_{\rm V}/E_{B-V} = $3.1 law and the reddening amounts
to $E_{B-V}$=0.87$\pm$0.01.

	\subsection{Distance}

   \begin{figure}
   \centerline{\psfig{file=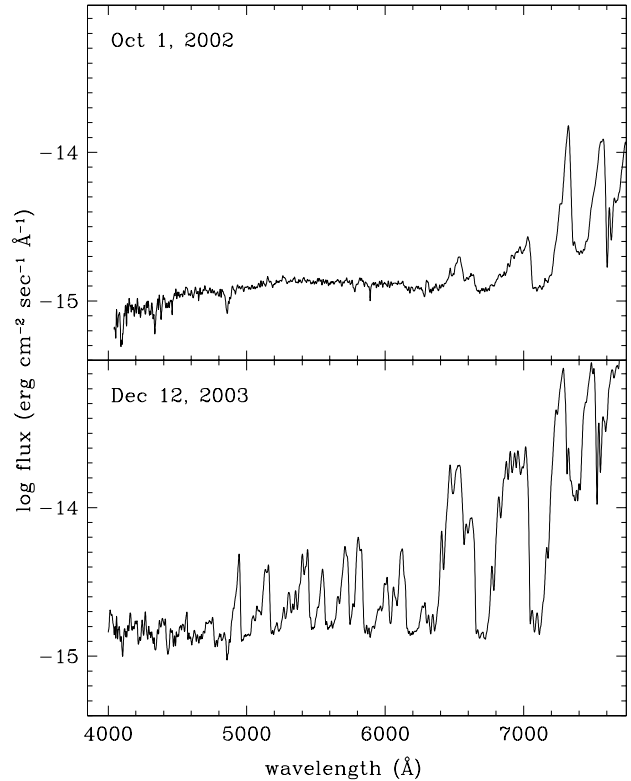,width=8.2cm}}
   \caption[]{Comparison between Asiago AFOSC spectra of V838~Mon taken in
   October 2002 and December 2003. At the earlier date the outbursting
   component is only visible toward the red, revealing the presence of a
   B3~V companion whose emission dominates the $U$, $B$ and $V$ bands at
   that time. The later spectrum illustrates how the emission of the
   outbursting component has retraced back to shorter wavelengths during the
   2003/04 visibility season, a change already described in its early
   development by Wagner et al. (2003).}
   \label{B3V}
   \end{figure}

There are several and independent arguments and evidences that support
a long distance scale to V838~Mon, of the order of 10~kpc.

Analysis of the high-resolution HST polarimetry images of the light-echo
led Bond et al. (2003) to place a lower limit of 6~kpc to the distance of
V838~Mon. Working on the same HST material, Tylenda (2004) revised the 
distance to 8$\pm$2~kpc.

The spectro-photometric parallax to the B3V companion derived in sect. 5 is
10~kpc. Furthermore, reversing the argument of the previous
section, we can enter Table~2 or Figure~5 with the reddening $E_{B-V}$=0.87
(corresponding to $A_V$=2.7) determined from the interstellar lines and the
colors of the B3V companion, obtaining the same distance of 10~kpc to V838~Mon.

Two other arguments are in favor of a large distance to V838~Mon, given its
proximity to the galactic plane ($b$=$-3^\circ$51$^\prime$ galactic
latitude). First, we have seen above that the HI radio observations in the
direction of V838~Mon reveal three components whose velocity match exactly
that of the three components seen in the interstellar absorption lines.
Given the fact that the radio observations integrate along the whole line of
sight through the Galaxy and that no HI is seen {\em beyond} V838~Mon (in
the sense that no corresponding interstellar lines are detected in the
high-resolution spectra), it is straightforward to conclude that V838~Mon
lies at great distance with most of the galactic gas, in that direction, in
front of it. Second, Schlegel et al. (1998) extinction maps are based on the
amount of infrared emission by the dust integrated along the whole line of
sight through the Galaxy. The amount of extinction affecting V838~Mon
($E_{B-V}$=0.87 corresponding to $A_V$=2.7 mag) match the value of Schegel
et al. maps in that direction, supporting again the notion that V838~Mon
lies at large galactocentric distances.

All these independent determinations support the conclusion that
V838~Mon lies in the outer part of the disk of the Galaxy, at a distance of
$\sim$10~kpc from the Sun, corresponding to a galacto-centric distance of
$\sim$17.5~kpc and a height above the galactic plane of $\sim$650~pc.

Tylenda (2004) labeled as ``naive interpretation'' the early distance
estimates of Munari et al. (2002a) and Kimeswenger et al. (2002) based on
the first determinations of the angular expansion rate of the light-echo on
early ground-based discovery images. The Munari et al. (2002a) implicit (but
quite obvious) assumption was that the light-echo was originating in a
circumstellar disk seen pole-on. Such an assumption was based on the fact
that the NaI and KI lines were not tracing a circumstellar component,
something to be expected in the case of a homogeneous spherical distribution
of material centered on the object itself.  It was only much later that high
resolution HST imaging revealed that there is a clear void of circumstellar
material precisely along the line of sight to V838~Mon, regardless of the
true 3D shape of the circumstellar dust giving rise to the light-echo. So,
the now ``naive'' approach was reasonable at the time.

	\subsection{Progenitor}

   \begin{figure}
   \centerline{\psfig{file=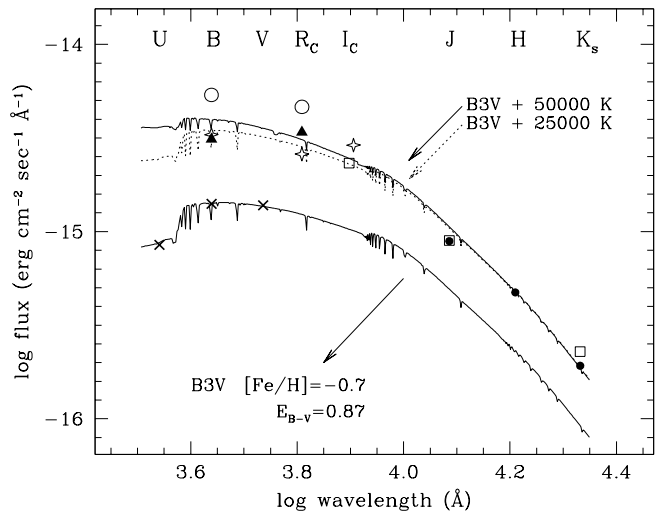,width=8.6cm}}
   \caption[]{The crosses represent the energy distribution of V838~Mon at the
   time of naked visibility of the hot companion ($V$=16.05, {\em B--V}=+0.68,
   {\em U--B}=$-$0.06) compared with a Kurucz's spectrum for a metal
   poor B3V star ($T_{\rm eff}$=19000~K, $\lg g$=4.0, [Fe/H]=$-$0.7)
   reddened by $E_{B-V}$=+0.87. The other symbols give the energy distribution
   of V838~Mon in quiescence: stars from Goranskii et al. (2004), triangles from
   Kimeswenger et al. (2002), squares from DENIS survey, dots from 2MASS survey,
   circles from USNO-B magnitudes re-calibrated against the comparison sequence
   of Munari et al. (2002), giving $B$=15.28 and $R_{\rm C}$=14.22. The quiescence
   energy distribution is fitted with the combination of Kurucz's spectra for the 
   B3V stars and the hotter companion.}
   \end{figure}

   \begin{figure}
   \centerline{\psfig{file=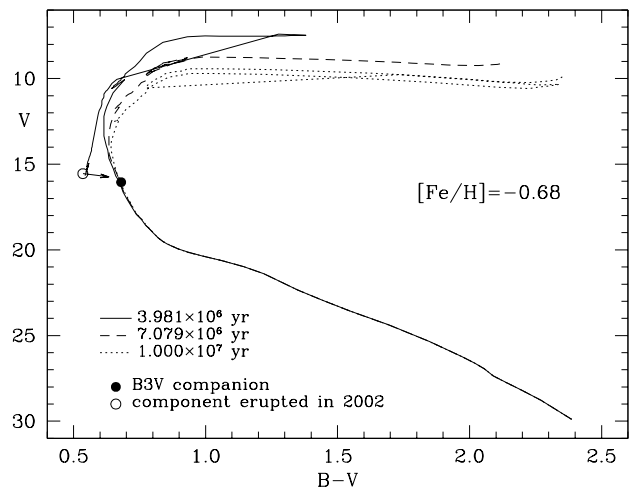,width=8.2cm}}
   \caption[]{Padova isochrones (from Bressan et al. 1993, Fagotto et al.
              1994, Girardi et al. 2000) for Z=0.004 and three
              ages scaled to the distance (10 kpc) and reddening ($E_{\rm
              B-V}$=0.87) of V838~Mon. The isochrones include the effect 
              of mass loss and are corrected for the
              reddening-induced distortion described in Fiorucci \& Munari
              (2003, their Figure 8). The open circle indicates the position
              of the V838~Mon component that erupted in 2002 if its
              temperature in quiescence was 50\,000~K, and the arrow gives
              the shift if the temperature was 25\,000~K (cf. the energy
              distribution fit in Figure~10).}
   \end{figure}

Figure~10 shows the energy distribution of the B3\,V companion in the {\em
UBV} bands ($<$$V$$>$=16.052, $<$$B$$>$=16.736, $<$$U$$>$=16.676 average values
from USNO 1.0m photometry described in sect.1) compared with a Kurucz's
synthetic spectrum (from the library of Munari et al. 2004) with parameters
appropriate for the B3\,V companion ($T_{\rm eff}$=19\,000~K, $\log g$=4.0,
[Fe/H]=$-$0.7). The match is excellent. In the same figure, the pre-outburst
brightness of V838~Mon from various sources are plotted. In the infrared
they come from the 2MASS and DENIS surveys, while the $B$ and $R_{\rm C}$
values are estimates of the same POSS-I and SERC plates according to
different authors that used different calibrations. The entries at $I$ band
(note the slightly different effective wavelengths) come from the DENIS
and POSS-II surveys.

The figure clearly shows that {\em the progenitor of the outbursting
component was brighter and hotter than the B3\,V companion}. By exactly how
much it depends on the still uncertain magnitude of V838~Mon on pre-outburst
photographic plates. While the brightness seems well established in the
infrared by the excellent agreement of DENIS and 2MASS data, it presents a
large scatter in the estimates by different authors of the same POSS and
SERC plates. It is clear from Figure~10 that a careful re-evaluation of all
existing pre-outburst photographic material would be beneficial in
establishing the nature of V838~Mon prior to the outburst. Such a detailed
analysis must first provide a perfect comparison sequence, a tight control
over the plate measurement and its modeling, a careful knowledge of the
exact band-pass profile of the given emulsion-filter-telescope-atmosphere
combination and plate developing conditions, as well as careful application
of color equation corrections.

The best fit to the V838~Mon quiescence data in Figure~10 is the B3\,V plus
a 50\,000~K star with $V$=15.55 and $B-V$=+0.535. Both stars are equally
reddened by $E_{B-V}$=0.87. The combined spectrum of the two stars is
plotted in Figure~10 as a solid line. Ignoring our re-estimate of the
magnitude of V838~Mon on the POSS and SERC plates listed in the caption of
Figure~10, a somewhat lower temperature for the eruption progenitor can be
obtained, with a lower limit of 25\,000~K. The combination of the spectra of
the B3\,V plus a 25\,000~K star with $V$=15.74 and $B-V$=+0.634 is plotted as
a dotted line in Figure~10.

The location of the components of the binary system V838~Mon is compared with
theoretical isochrones in Figure~11, for the [Fe/H]=$-$0.7 metallicity
appropriate to the galactocentric distance of V838~Mon and for the solar
case as a matter of reference. The Padova theoretical isochrones have been
scaled to a distance of 10~kpc and $E_{B-V}$=0.87 reddening following the
standard $R_V$=3.1 extinction law. The reddening dependent deformation over
the HR plane discussed and quantified by Fiorucci \& Munari (2003, cf.
their Figure~8) for the {\em UBVR$_{\rm C}$I$_{\rm C}$} photometric system
(Vilnius and USA reconstructions) has been applied to the isochrones in Figure~11.

As displayed by Figure~11, there is a perfect match of the [Fe/H]=$-$0.7
isochrones with the observed magnitude and colors of the B3\,V component,
with the progenitor of the outbursting companion lying on the isochrone for
an age of 4 million yr, close to the position of the central ignition of Carbon 
for an initial mass of $\sim$65~M$_\odot$. The progenitor lies close to  
the region where Wolf-Rayet stars are usually located. It would be harder 
to fit with theoretical evolutionary tracks the position of the progenitor 
of the outbursting component if its temperature would be decreased to the 
lower limit of 25\,000~K of Figure~10, indicated by the head of the arrow 
in Figure~11.

We therefore conclude that the progenitor of the outbursting component of
the binary system V838~Mon was hotter, brighter and born far more massive
than the 7~M$_\odot$ B3\,V companion, of an age not far from 4 million yr. 
In this respect, the circumstellar scattering material giving rise to the
light-echo is probably the result of the massive mass loss that such massive
objects experience, where $\sim$half of the mass is lost during the main
sequence phase. Given the location of V838~Mon in the outskirts of the
Galactic disk at galactic longitude $l$=218$^\circ$, the large mass and
young age could sound problematic. However, young clusters and massive stars
at great distances in the anti-center direction are already known. For ex.,
Fitzgerald \& Moffat (1976) reported a distance of $\sim$7~kpc for the very
young cluster Ruprecht~44 ($l$=245$^\circ$) rich in O and WR stars, or Marco
et al. (2001) derived a distance of 6~kpc and an age of 4 Myr for the
cluster NGC 1893 ($l$=173$^\circ$) that harbors O5V member stars (which
mass is $\sim$65~M$_\odot$ according to Strai\v{z}ys \& Kurilien\'e (1981)
tabulation).

The outburst experienced in 2002 does not appear as the terminal event in
the life of the massive progenitor, but instead more probably as a
thermonuclear shell flash in the outer layers of the star as could be
expected in the case of He after most of the H-rich outer envelope has been
blown away by the strong wind that characterize this type of stars. A
detailed analysis of the evolutionary status of the progenitor will be
presented elsewhere (Munari et al., in preparation).

\acknowledgements{This work has been supported in part by Italian
COFIN-2002 grant, Polish KBN Grant No. 2 P03D 019 25, and NSF and NASA
grants to ASU.  We would like to thank F.d'Antona and O.Straniero for
useful comments.}

\clearpage

\setcounter{table}{2}

   \begin{table}[t]
   \centering
   \caption[]{Journal of high-resolution spectroscopic observations. {\em
   HUT} = heliocentric UT date (dd/mm/yy); {\em HJD} = corresponding
   heliocentric JD ($-$2452000); {\em N} = number of spectra; {\em Exp$_{\rm
   T}$}= total exposure time (in sec); {\em S/N} = signal-to-noise of the
   continuum around NaI and/or KI in the combined spectrum; {\em D} =
   dispersion; {\em Res} = resolving power ($\lambda/\Delta\lambda$); {\em
   slit} = aperture on the sky (in arcsec) of the spectrograph entrance
   slit; {\em lines} = interstellar lines observed.}
   \centerline{\psfig{file=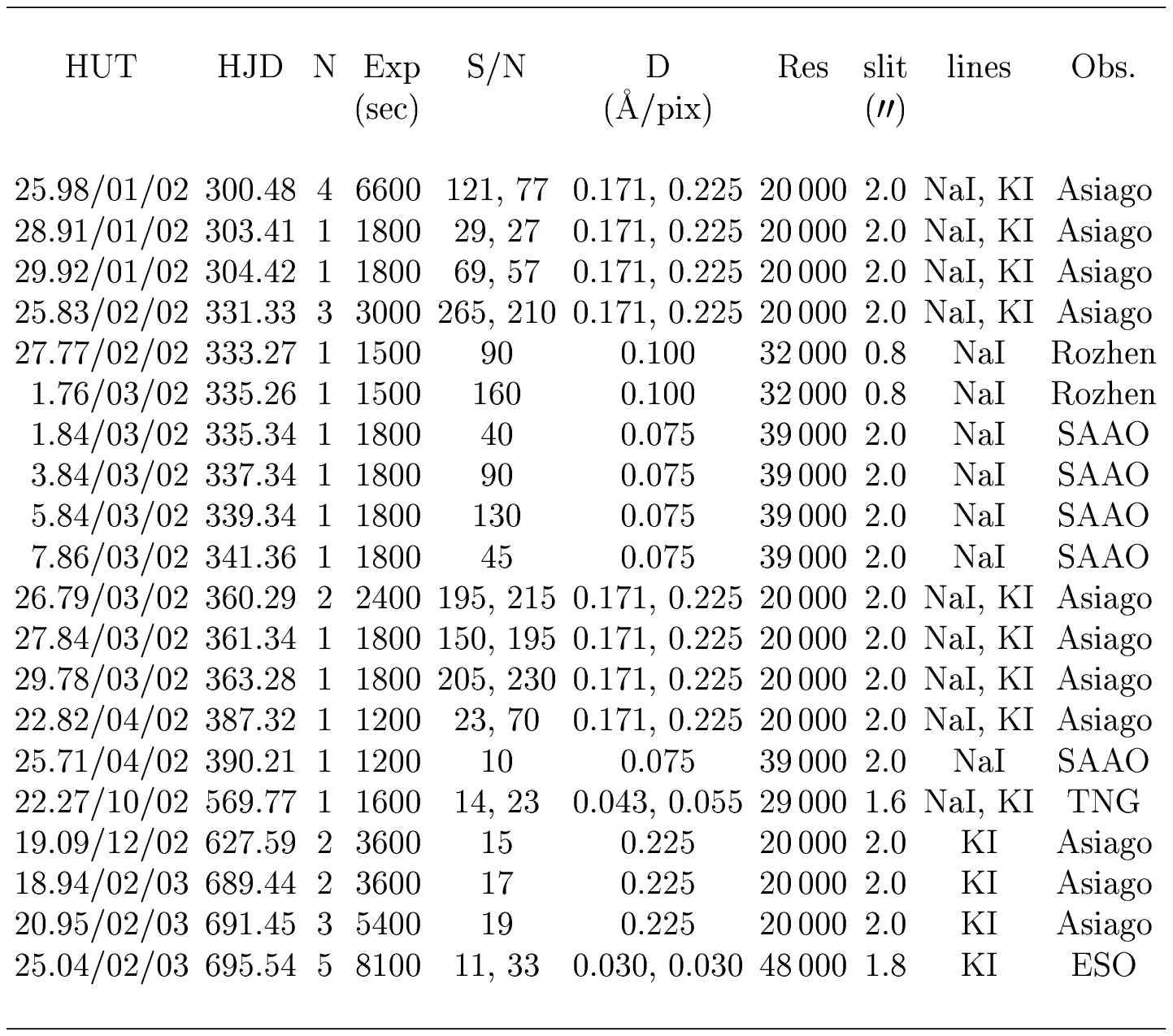,width=8.8 cm}}
   \label{Tab_1}
   \end{table}

   \begin{table*}
   \centering
   \caption[]{Results of the multi-gaussian fitting to the NaI and KI
   interstellar lines in the high resolution spectra of V838~Mon listed in
   Table~1 (rest wavelengths at 5895.923, 5889.953 and 7698.979~\AA\ for
   NaI~D1, D2 and KI, respectively). All wavelengths ($\lambda$) are
   heliocentric corrected, and the FWHMs ($\Gamma$) are corrected for the
   width of the instrumental PSF.  Both are expressed in km~sec$^{-1}$. The
   first two digits of the wavelengths (59 for both NaI D1 and D2, 76 or 77
   for KI) are dropped for table compactness. The equivalent widths are in
   \AA, and computed relative to a linear fit of the stellar continuum
   adjacent to the lines.  The nomenclature {\em blue} and {\em red}
   components refers to the broader line at $<$+37.7$>$ and the narrower one
   at $<$65.1$>$ km~sec$^{-1}$, respectively, that compose the profile of
   all interstellar lines, and that are well illustrated by the NaI~D1 line 
   at the top of Figure~1 (superimposed on the wide NaI P-Cyg stellar profile).}
   \centerline{\psfig{file=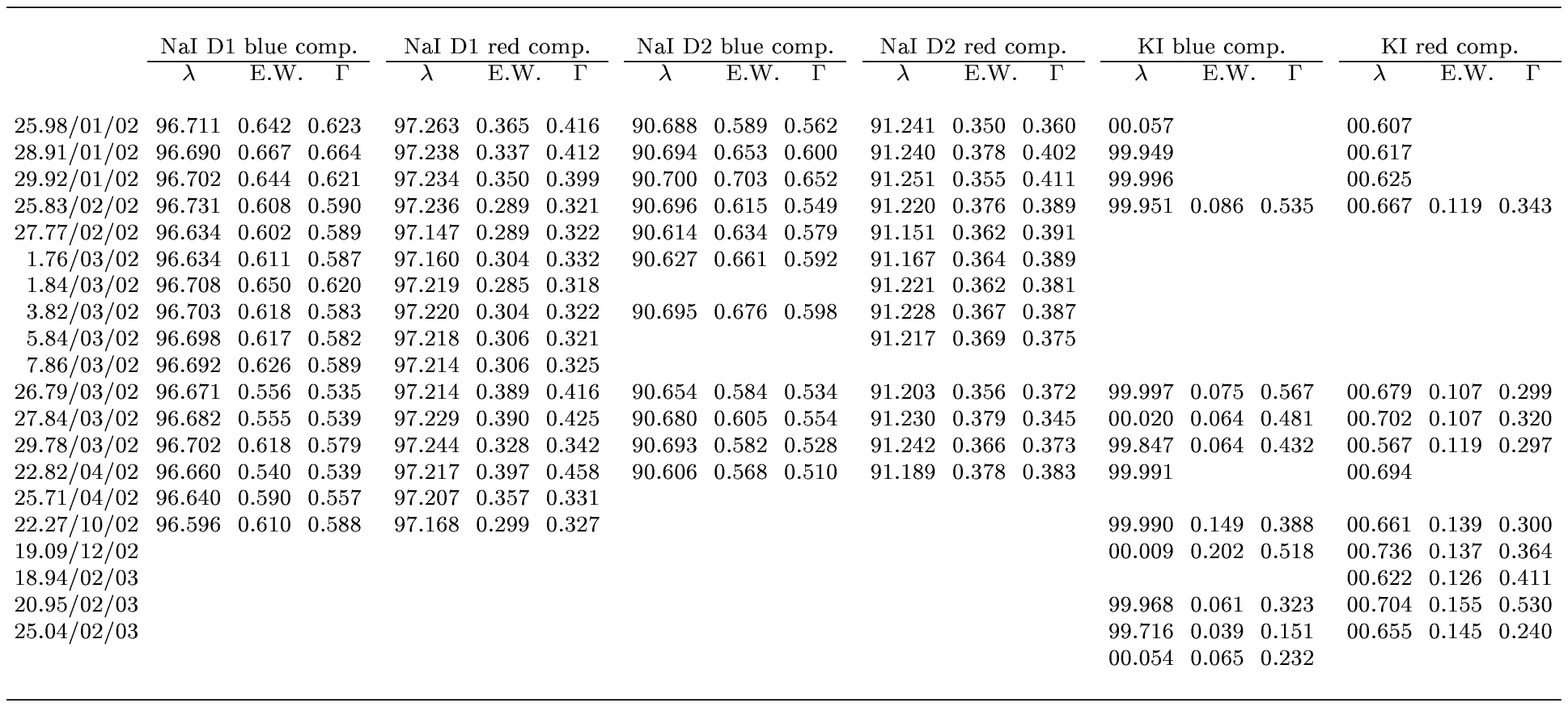,width=18.3 cm}}
   \label{Tab_2}
   \end{table*}

\end{document}